\documentclass{elsart}

\usepackage{natbib}

\usepackage{graphicx}

\usepackage{amssymb}
\usepackage{amsmath}

\usepackage{lineno}

\usepackage{color}

\newcommand{\code}[1]{\textsc{#1}}
\newcommand{\incl}{\code{INCL4}}
\newcommand{\isabel}{\code{Isabel}}
\newcommand{\geminipp}{\code{GEMINI++}}
\newcommand{\abla}{\code{ABLA07}}
\newcommand{\dresner}{\code{Dresner}}
\newcommand{\gem}{\code{GEM}}
\newcommand{\smm}{\code{SMM}}
\newcommand{\mcnpx}{\code{MCNPX}}
\newcommand{\geant}{\code{Geant4}}
\newcommand{\phits}{\code{PHITS}}

\begin{document}

\begin{frontmatter}


\title{Influence of nuclear de-excitation on observables relevant for space exploration}

\author[liege]{Davide Mancusi\corauthref{cor}}\corauth[cor]{Corresponding author.}\ead{d.mancusi@ulg.ac.be},
\author[cea]{Alain Boudard},
\author[liege]{Joseph Cugnon},
\author[cea]{Jean-Christophe David},
\author[cea]{Sylvie Leray}

\address[liege]{University of Li\`ege, AGO Department, all\'ee du 6 Ao\^ut 17, b\^at. B5\\B-4000 Li\`ege 1, Belgium}
\address[cea]{CEA/Saclay, Irfu/SPhN, 91191 Gif-sur-Yvette, Cedex, France}

\begin{abstract}

  The composition of the space radiation environment inside spacecrafts is modified by the interaction with shielding
  material, with equipment and even with the astronauts' bodies. Accurate quantitative estimates of the effects of nuclear
  reactions are necessary, for example, for dose estimation and prediction of single-event-upset rates. To this end, it is
  necessary to construct predictive models for nuclear reactions, which usually consist of an intranuclear-cascade or
  quantum-molecular-dynamics stage, followed by a nuclear-de-excitation stage.

  While it is generally acknowledged that it is necessary to accurately simulate the first reaction stage, transport-code
  users often neglect or underestimate the importance of the choice of the de-excitation code. The purpose of this work is
  to prove that the de-excitation model is in fact a non-negligible source of uncertainty for the prediction of several
  observables of crucial importance for space applications. For some particular observables, the systematic uncertainty
  due to the de-excitation model actually dominates the total uncertainty. Our point will be illustrated by making use of
  nucleon-nucleus calculations performed with several intranuclear-cascade/de-excitation models, such as the Li\`{e}ge
  Intranuclear Cascade model (\incl) and \isabel\ (for the cascade part) and \abla, \dresner, \gem, \geminipp\ and \smm\
  (on the de-excitation side).

\end{abstract}


\end{frontmatter}

\parindent=0.5 cm

\section{Introduction}

Radiation exposure is generally recognised as one of the major hazards for the safety of spacecraft crews and equipment on
long-term interplanetary missions. The most obvious countermeasure to date consists in shielding spacecraft and space
stations with suitable materials, to reduce the dose absorbed by astronauts and thus the risk. However, the interaction of
the space radiation environment with the shielding material generates an intense secondary radiation field, whose
properties must be accurately assessed in order to make reliable risk predictions.

The secondary radiation field is typically due to proton-nucleus or nucleus-nucleus reactions at energies between
a few hundred $A$MeV and several $A$GeV. Proton-nucleus reactions are especially frequent because of the high flux of Galactic
Cosmic Ray (GCR) protons (87\% of the total GCR baryonic flux) \citep{durante-radiation}. Thus, the estimation of the
quality of the secondary radiation field crucially depends on the availability of accurate nucleon-nucleus reaction models
in the 0.1--10~GeV incident-energy range.

It is generally agreed that such reactions proceed in two stages: a fast ($\sim10^{-22}$~s) dynamical stage generates an
excited compound nucleus, which subsequently de-excites by emitting nucleons or light ions on a much longer time scale
($\sim10^{-12}$~s). This distinction mirrors the different theoretical approaches that have been most successful in
modelling the two stages. The fast stage of nucleon-induced reactions is traditionally described by intranuclear-cascade
(INC) models \citep{serber-reactions}, sometimes followed by an intermediate pre-equilibrium step; however, several other
models (e.g.\ Quantum Molecolar Dynamics, Boltzmann-Uehling-Uhlenbeck, Vlasov-Uehling-Uhlenbeck) are often considered as
well. The de-excitation of the compound-nucleus is usually handled by statistical decay or break-up models.

Transport-code users are generally aware of the great influence of the choice of the dynamical model on the calculation
results. However, several observables are actually even more sensitive to the choice of the de-excitation model, which is
generally just taken for granted. The goal of this work is to raise awareness about the importance of the choice of the
de-excitation model for the estimation of several quantities relevant for radioprotection in space. We demonstrate that
the systematic uncertainty due to the de-excitation stage can be as large as or larger than the uncertainty connected with
the modelling of the dynamical stage.

To prove this point, we have performed thin-target nucleon-nucleus calculations with two INC models (\incl, \isabel)
coupled to five de-excitation codes (\abla, \geminipp, \gem, \smm, \dresner). Thin-target calculations neglect transport
of the reaction products in the surrounding material. Thick-target calculations would be better suited to assess the
influence of de-excitation models. However, none of the major transport codes presently available offers a vast choice of
nuclear-de-excitation models, which is indispensable for such a sensitivity study. On the other hand, the choice of
nucleon-nucleus reactions is mainly motivated by the large cosmic-ray proton flux and by the abundance of hydrogen-rich
materials in shielding. Moreover, the uncertainty connected with the dynamical stage is smaller for nucleon-nucleus than
for nucleus-nucleus reactions, which can be affected by important collective phenomena such as bounce-off and
squeeze-out. Thus, nucleon-nucleus reactions represent the best scenario to highlight the importance of de-excitation.

\section{Model description}

We will limit ourselves to give brief descriptions of the salient features of each model. The reader is referred to the
relevant literature for further details.

\subsection{Intranuclear-cascade models}

The \textbf{\isabel} model \citep{yariv-isabel1,yariv-isabel2} is a time-like intranuclear-cascade model that has widely
contributed to the understanding of nucleon-nucleus and nucleus-nucleus collisions in the previous decades. It considers
the target nucleus as a continuous medium (Fermi sea), which is perturbed by the collisions induced by the incoming
cascading particles. \isabel\ features one of the most detailed treatments of the $\Delta(1232)$ resonance.

The Li\`ege Intranuclear-Cascade model (\textbf{\incl}) \citep{boudard-incl,cugnon-incl45_nd2010} is a modern addition to the INC
family. Its main feature is that it represents the target nucleus as a collection of quasi-particles moving along straight
trajectories in a potential well. It contains refined physics inspired by the optical-model phenomenology (e.g.\ energy-
and isospin-dependent potentials for nucleons and pions), as well as a dynamical coalescence mechanism that makes the
model applicable to the study of the observed emission of fast composite particles like deuterons, tritons and heavier
nuclei. The $\Delta(1232)$ resonance is also treated in detail.

Both models can be applied to roughly the same incident-energy range (150~MeV--3~GeV).

\subsection{De-excitation models}

\textbf{\abla} \citep{ricciardi-abla_vienna} is the latest version of the de-excitation code \code{ABLA}, originally
developed by K.-H. Schmidt and his team \citep{gaimard-abla}. While the \abla\ model is well known for its accurate
treatment of fission, it has recently been extended to describe evaporation of intermediate-mass fragments (IMFs),
multifragmentation and to include competition between particle and gamma-ray emission. An older version of the
\incl+\code{ABLA}\ cascade+de-excitation model is included in \mcnpx\ and \geant.

The \textbf{\dresner} model \citep{dresner-model} is the default de-excitation option for the \mcnpx\ transport code. It
contains a simple Weisskopf-Ewing evaporation formalism, a simple Bohr-Wheeler fission model and utilises a Fermi break-up model to describe the de-excitation
of light nuclei. It is nowadays unmaintained.

The Generalized Evaporation Model \citep[\textbf{\gem},][]{furihata-gem} is a Japanese fission-evaporation code and the
default option of the \phits\ transport code \citep{iwase-phits}. Its uses the Weisskopf-Ewing evaporation formalism to
describe the emission of nucleons and nuclei up to Mg.

The \textbf{\geminipp} model, developed by R.\ Charity \citep{charity-gemini++}, is a binary de-excitation code, which is
mostly notable for its fission-like description of IMF emission. Since it was originally developed to describe
nucleus-nucleus reactions, \geminipp\ features an accurate treatment of orbital and intrinsic angular momentum. It is
actually the only de-excitation model employed in this work that uses the Hauser-Feshbach evaporation formalism.

Finally, the Statistical Multifragmentation Model \citep[\textbf{\smm},][]{bondorf-multifragmentation} is an actively
developed code which combines the compound-nucleus de-excitation processes at low energy and simultaneous break-up at high
energy. It has largely been used to study the behaviour and properties of nuclear matter at subnuclear density and it has
been applied to multifragmentation reactions in astrophysics.

\section{Results}

We will now separately consider the influence of the choice of the de-excitation model on several observables of interest
for radioprotection in space.

\subsection{Charge-changing cross sections}

Since shielding materials are often rich in hydrogen, GCR nuclei frequently undergo nuclear reactions with target
protons. Such reactions can easily be modelled in inverse kinematics (i.e.\ as proton-induced reactions) in the framework
of cascade+de-excitation models. The charge-changing cross section for the GCR projectile is thus given by the
corresponding cross section for the target nucleus in inverse kinematics.

\begin{figure}
  \centering
  \includegraphics[width=0.85\linewidth]{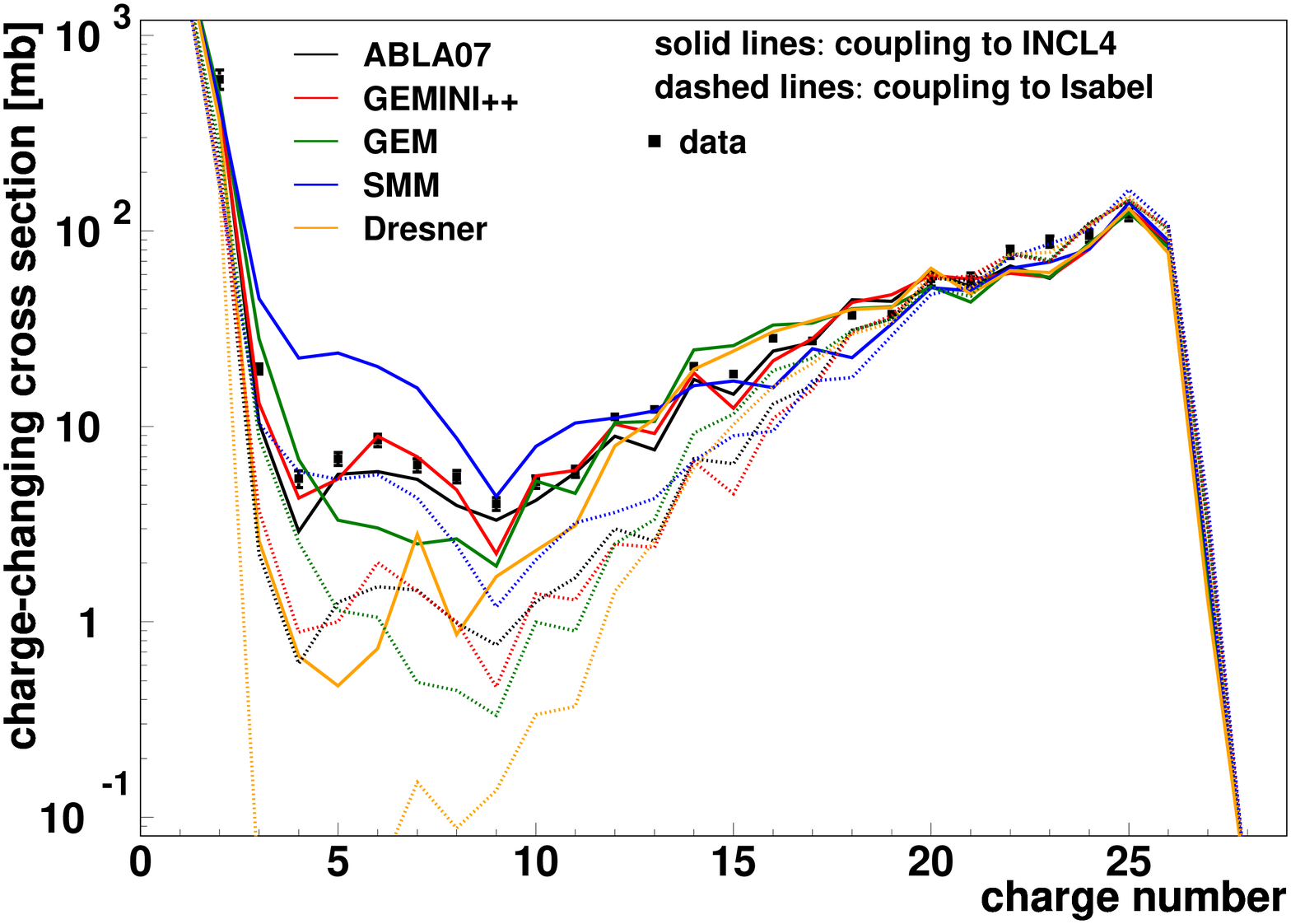}
  \caption{Charge-changing cross sections for 1-GeV \textit{p}+$^\text{56}$Fe. Cascade and de-excitation models are
    indicated by the dashing and the colouring, respectively. Experimental data from
    \cite{napolitani-fe,villagrasa-fe,legentil-fe}.}
\label{fg:ccxs}
\end{figure}

Fig.~\ref{fg:ccxs} shows a comparison of charge-changing cross sections for 1-GeV \textit{p}+$^\text{56}$Fe. The reaction
has been selected because cosmic iron nuclei are responsible for a large fraction of the GCR equivalent dose; their
charge-changing cross section thus need to be accurately predicted. The dashing style of the lines identifies the cascade
model. The uncertainty connected with the choice of the de-excitation model can thus be estimated by comparing lines with
the same dashing style. We observe that the spread is largest for the lightest fragments (IMFs). In the $3\lesssim
Z\lesssim15$ charge range, the uncertainties connected with cascade and de-excitation are comparable. This can be
explained by observing that this region is populated by cascade events with high excitation energies (several hundred
MeV), which thus amplify the sensitivity to the details of the de-excitation model. At the same time, since events with
high excitation energies are quite rare, their cross sections and distributions (and, thus, the $3\lesssim Z\lesssim15$
cross sections) are also sensitive to the choice of the cascade model.

\subsection{Dose from secondary neutrons}

\begin{figure}
  \centering
  \includegraphics[width=0.85\linewidth]{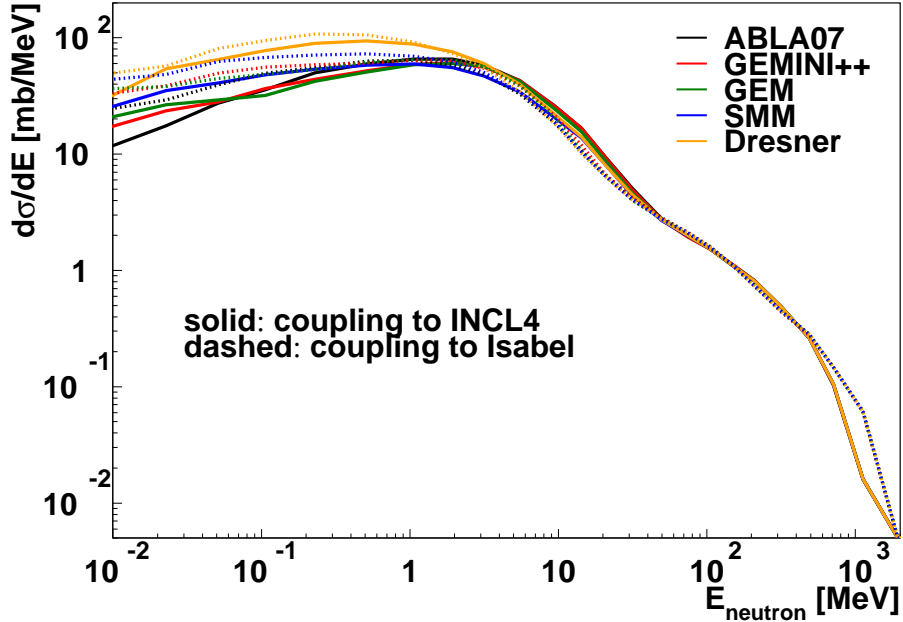}
  \caption{Energy-differential neutron-production cross sections, for the 1-GeV \textit{p}+$^\text{27}$Al
    reaction. Cascade and de-excitation models are indicated by the dashing and the colouring, respectively.}
\label{fg:nxsec}
\end{figure}

We proceed by illustrating the influence of de-excitation models on secondary neutron production. Fig.~\ref{fg:nxsec}
shows the calculated energy-differential neutron-production cross section for the 1-GeV \textit{p}+$^\text{27}$Al
reaction, which is taken as a representative of the reactions responsible for the creation of the strong secondary-neutron
background in space vessels. It appears that de-excitation models mostly affect the low end of the spectrum, up to about
10--20~MeV. Above this energy, the cross section is entirely determined by the cascade stage.

Fig.~\ref{fg:nxsec} by itself is not conclusive about the importance of de-excitation models for the estimation of
secondary-neutron doses. An accurate assessment would require a full three-dimensional transport calculation in a
realistic geometry and with a realistic source. This is beyond the scope of this paper. We can nevertheless make a rough
evaluation by using energy-dependent dose-conversion coefficients, such as those calculated by \cite{sato-coeff} using the
\phits\ code and the ICRP/ICRU reference phantoms. Hence, we suppose that the ICRP/ICRU reference male phantom is
irradiated isotropically by the neutrons fields depicted in Fig.~\ref{fg:nxsec}. By multiplying the energy-differential
cross section by Sato \etal's conversion coefficients and integrating over the neutron energy, we can obtain an estimate
of the effective dose to the human body.

\begin{figure}
  \centering
  \includegraphics[width=0.85\linewidth]{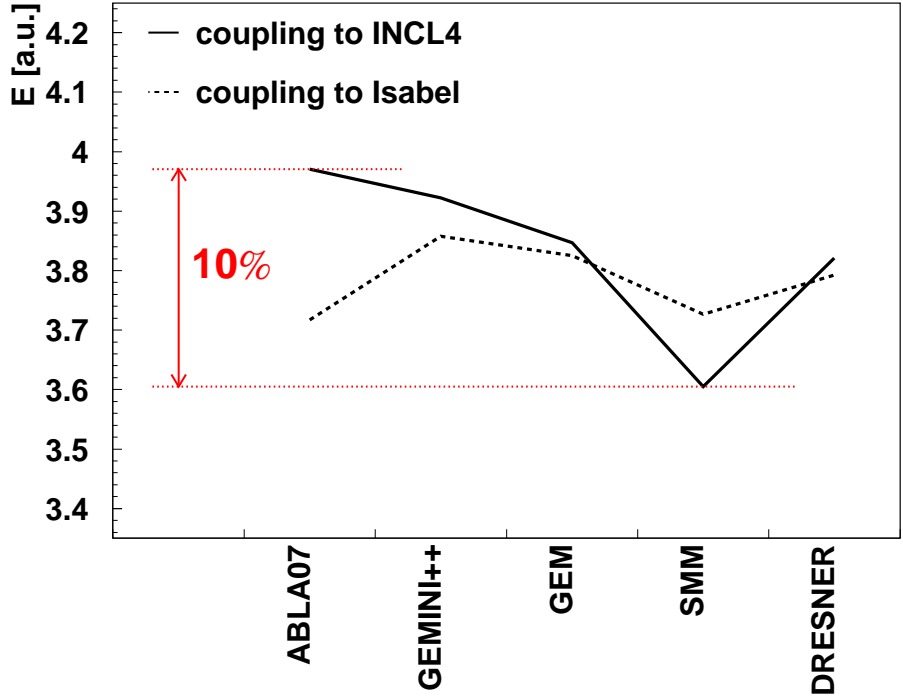}
  \caption{Effective dose delivered by secondary neutrons of the 1-GeV \textit{p}+$^\text{27}$Al reaction to the ICRP/ICRU
    male reference phantom, estimated using \protect\citeauthor{sato-coeff}'s \protect\citeyearpar{sato-coeff}
    fluence-to-dose conversion coefficients for the isotropic irradiation geometry.}
\label{fg:ndose}
\end{figure}

Fig.~\ref{fg:ndose} shows the result of this procedure. Since the fluence-to-dose coefficients are small at low energy,
the uncertainty connected with de-excitation is suppressed, but it is still of the same order of magnitude as the
uncertainty due to cascade. Neither of them anyway exceeds 10\%.

\subsection{Dose from secondary protons}

The results are radically different for secondary protons. It has been shown \citep{mancusi-1gevprotons} that secondary
protons can be responsible for a large increase in dose deposition (as much as 60\%) after thick shields. This phenomenon
is due to the large low-energy, high-LET fluxes of secondary protons produced in high-energy collisions.

\begin{figure}
  \centering
  \includegraphics[width=0.85\linewidth]{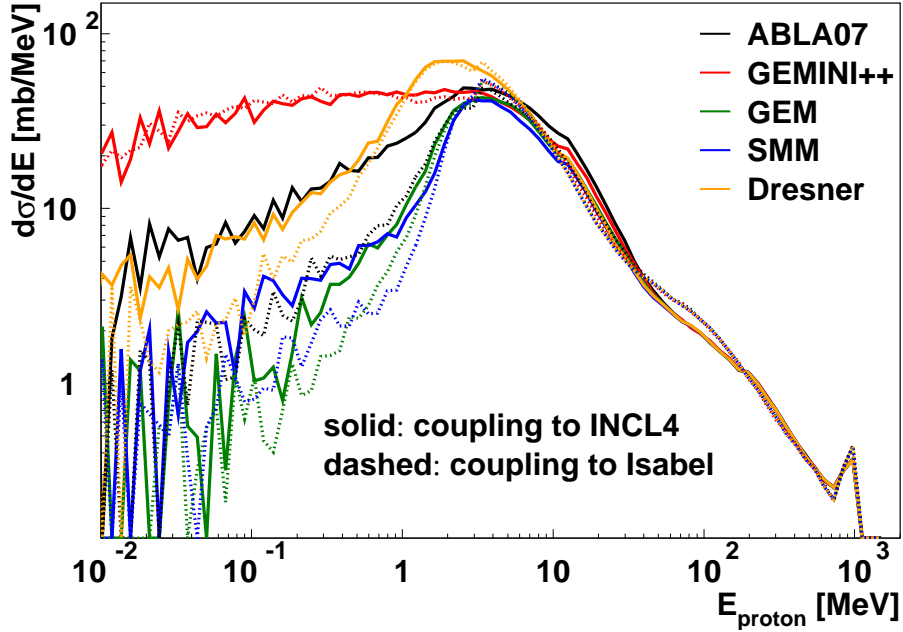}
  \caption{Energy-differential proton-production cross sections, for the 1-GeV \textit{p}+$^\text{27}$Al
    reaction. Cascade and de-excitation models are indicated by the dashing and the colouring, respectively.}
  \label{fg:pxsec}
\end{figure}

We have again chosen 1-GeV \textit{p}+$^\text{27}$Al as a representative reaction between space radiation and shielding
material. The energy-differential cross sections for proton production are shown in Fig.~\ref{fg:pxsec}. As in the case of
neutrons, de-excitation yields dominate in the low-energy end of the spectrum. The uncertainty however appears to be much
larger than for neutrons.

A rough estimate of the dose delivered by the depicted proton spectra can be obtained by assuming that protons deliver
dose proportionally to their LET. This is approximately true for thin biological targets (e.g.\ skin) placed close to the
shield. The approximation is less accurate for thick biological targets, where slowing-down and secondary reactions play
an important role. In such a case, one could use fluence-to-dose conversion coefficients, as it was done in the previous
section for neutrons.

\begin{figure}
  \centering
  \includegraphics[width=0.85\linewidth]{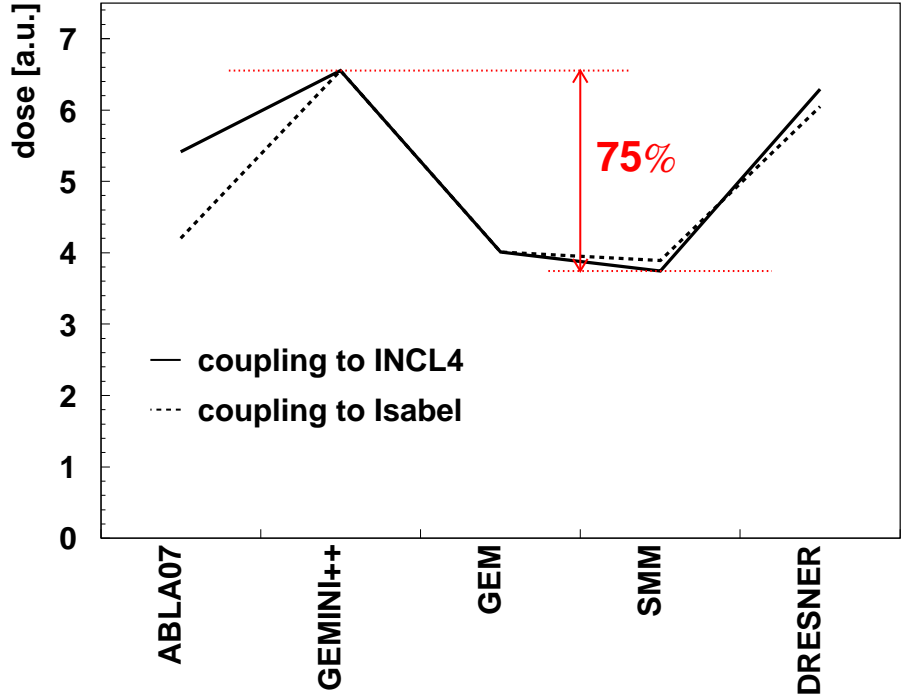}
  \caption{Dose delivered to a thin biological target by secondary protons of the 1-GeV \textit{p}+$^\text{27}$Al
    reaction.}
\label{fg:pdose}
\end{figure}

If however the biological target is assumed to be thin, one can multiply the spectra in Fig.~\ref{fg:pxsec} by the ICRU
reference LET-energy curves \citep{icru-report49} and integrate over the proton energy to obtain an estimate of the
delivered dose. Since LET is a decreasing function of energy above $\sim0.1$~MeV, the variance among the proton spectra
will be amplified, contrarily to secondary neutron doses. Fig.~\ref{fg:pdose} reveals that this result is found to be almost independent of the
cascade model; on the other hand, different de-excitation models can yield doses that can differ by as much as 75\%.

\subsection{Recoil-energy distributions}

We finally turn to recoil-velocity distributions, which are important for the prediction of Single-Event-Upset (SEU) rates
of electronic devices. SEUs are typically induced by nuclear reactions between secondary neutrons and silicon nuclei. Part
of the recoil kinetic energy acquired by the target fragment is lost to the ionisation of the silicon wafer, which may
upset the charge state of the electronic circuits therein contained. It is therefore crucial to be able to predict the
recoil-energy distribution of the target fragments.

\begin{figure}
  \centering
  \includegraphics[width=0.85\linewidth]{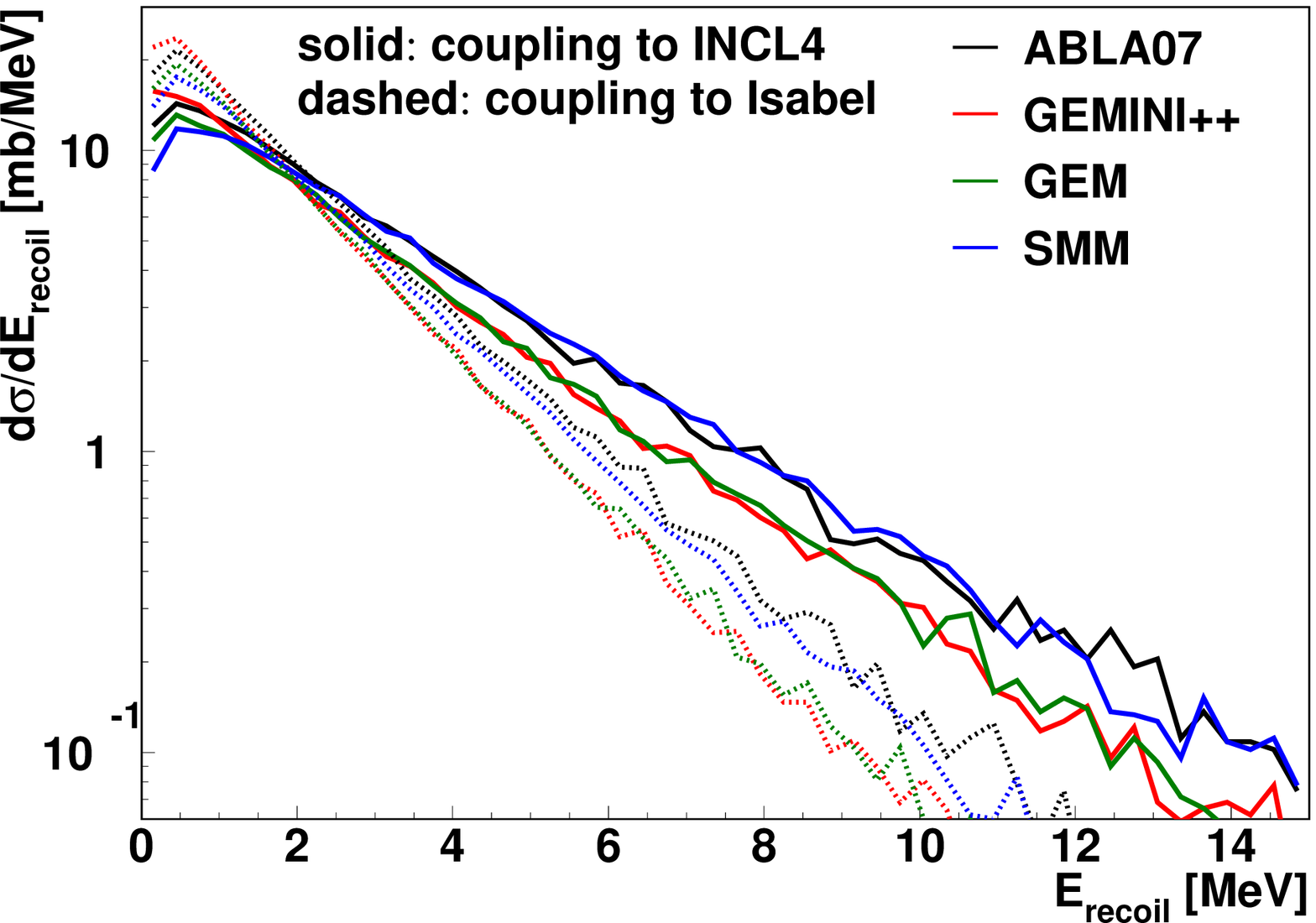}\\
  \includegraphics[width=0.85\linewidth]{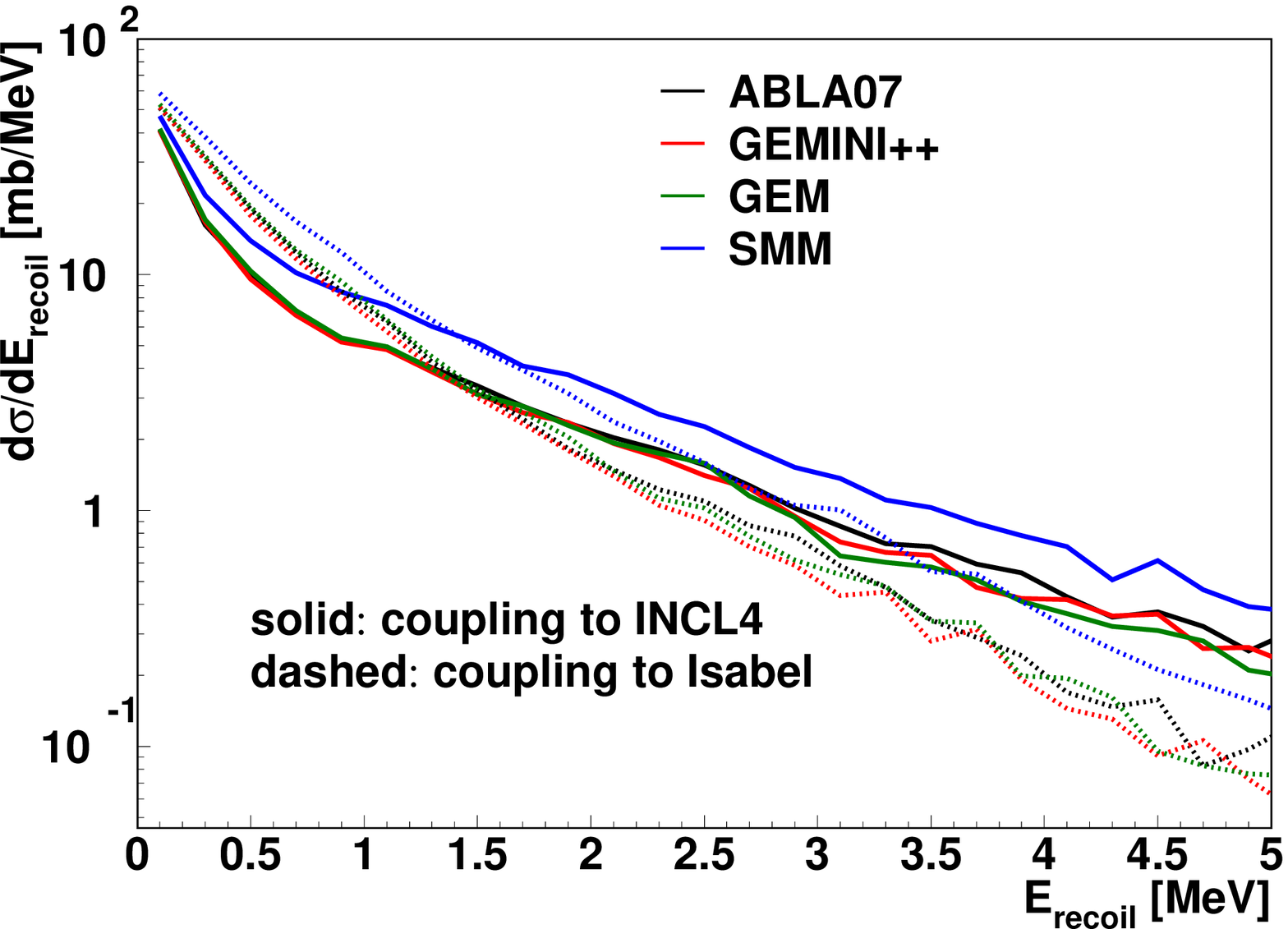}
  \caption{Recoil-energy distributions induced by the 200-MeV \textit{n}+$^\text{28}$Si reactions. Top panel: fragments
    with charge number $Z=11$, $12$. Bottom panel: $Z\geq13$. Cascade and de-excitation models are indicated by the
    dashing and the colouring, respectively.}
\label{fg:seu}
\end{figure}

We have chosen 200-MeV \textit{n}+$^\text{28}$Si as a representative reaction for the induction of SEUs. Fig.~\ref{fg:seu}
shows differential cross sections as functions of the recoil kinetic energy for different charges of the recoiling
fragments. Since the probability to induce a SEU increases with increasing LET of the particle, we focus here on the
nuclei with the highest charges (top panel: $Z=11$, $12$; bottom panel: $Z\geq13$). It is apparent that the main source of
uncertainty here is the cascade model. Curves of the same dashing style tend to cluster together, while different
de-excitation are only responsible for minor modulations. We thus conclude that recoil-velocity distributions for
fragments close to the target are rather insensitive to the choice of the de-excitation model.


\section{Conclusions and perspectives}

We have studied the influence of nuclear-de-excitation models on several observables of interest for radioprotection of
humans and equipment in space. The analysis was performed in the framework of nucleon-nucleus reactions, using two
different intranuclear-cascade models (\incl, \isabel) to simulate the dynamical stage of the reaction. Five different
de-excitation models (\abla, \geminipp, \gem, \smm, \dresner) were selected.

We have proved that the choice of the de-excitation model introduces a non-negligible uncertainty on charge-changing cross
sections for intermediate-mass fragments. Likewise, secondary-proton doses in thin biological targets can vary by as much
as 75\% with the choice of the de-excitation model. On the other hand, recoil-velocity distributions, which are essential
for the determination of single-event-upset rates, and secondary-neutron doses are less sensitive to the de-excitation
stage. One can summarise these results by observing that the observables that are sensitive to de-excitation are typical
of violent nuclear reactions. Very central collisions indeed lead to high excitation energies at the end of the cascade
stage, which amplify the sensitivity to the parameters and assumptions of the de-excitation models.

The influence of de-excitation models should be further investigated by performing thick-target calculations with
realistic geometries and radiation sources. However, this requires several de-excitation models to be available within the
same transport code. None of the major transport codes offer this possibility today. Hence, we encourage transport-code
developers to consider the inclusion of several de-excitation models.

Finally, it is less clear whether our conclusions would still hold in the context of nucleus-nucleus reactions, whose
importance in space is capital because of the large dose fraction delivered by GCR heavy ions. The strategy outlined in
the present paper can be applied to nucleus-nucleus reactions as well, provided that several dynamical reaction models are
available.

\end{document}